\documentclass[lettersize,journal]{IEEEtran}

\usepackage{cite}
\usepackage{amsmath,amssymb,amsfonts}
\usepackage{mathtools} % coinq command
\usepackage{algorithm}
\usepackage{algpseudocode} % some helper functions
\algrenewcommand\algorithmicrequire{\textbf{Input:}}
\algrenewcommand\algorithmicensure{\textbf{Output:}}

\usepackage{textcomp}

\usepackage{array}
\usepackage{stfloats}

\usepackage[hyphens]{url}
\usepackage[hidelinks]{hyperref}  % Internal linking between
\hypersetup{breaklinks=true}
\usepackage{verbatim}
\usepackage{multirow}
\usepackage{multicol}
\usepackage{enumitem}

\usepackage{graphicx, color}
\usepackage[caption=false]{subfig}
\usepackage{cite}
\usepackage{url}

\usepackage{acronym} % Allows the additiong of acronyms in the document
\usepackage{xcolor} 
\usepackage{tabularx}
\usepackage{gensymb} % add symbols such as °
\usepackage{booktabs}
\usepackage{lmodern} % Fonts

\begin{document}

\title{Toward Real-Time Digital Twins \\ of EM Environments: Computational Benchmark for Ray Launching Software}

\author{Michele Zhu, Lorenzo Cazzella, Francesco Linsalata, Maurizio Magarini, Matteo Matteucci, Umberto Spagnolini
\thanks{The work of Francesco Linsalata, Maurizio Magarini, and Umberto Spagnolini was partially supported by the European Union under the Italian National Recovery and Resilience Plan (NRRP) of NextGenerationEU, partnership on “Telecommunications of the Future” (PE00000001 - program “RESTART”, Structural Project 6GWINET).}
\thanks{Michele Zhu, Lorenzo Cazzella and Francesco Linsalata contributed equally to this research.}
\thanks{This work has been published in IEEE Open Journal of Communication Society. doi:
https://doi.org/10.1109/OJCOMS.2024.3463963}}

\maketitle

\begin{abstract}

Digital Twin has emerged as a promising paradigm for accurately representing wireless communication electromagnetic environments. The resulting virtual representation of reality facilitates comprehensive insights into the propagation environment, empowering multi-layer decision-making processes at the physical communication level.
This paper investigates the impact of ray-based model simulation within real-time Digital Twins.
A benchmark for ray-based propagation simulations is presented to evaluate computational time, considering two urban scenarios characterized by different mesh complexity, single and multiple wireless link configurations, and simulations with/without diffuse scattering. Exhaustive empirical analyses are performed showing the behavior of different ray-based solutions. By offering standardized simulations and scenarios, this work provides a technical benchmark for practitioners involved in the implementation of real-time Digital Twins and optimization of ray-based propagation models.
\end{abstract}


\section{Introduction}
\IEEEPARstart{T}{he} accurate modeling of electromagnetic (EM) environment has significant implications for improving key parameters in network design and deployment \cite{7509384}. Within this context, the concept of the Digital Twin (DT) emerges, representing a sophisticated but interesting approach to constructing precise digital replicas of physical entities or processes. These DTs provide real-time insight into their corresponding physical counterparts, effectively integrating simulations with existing databases \cite{DTMagazine}. This framework necessitates the development of appropriate models for the continuous update and enhancement of the virtual representations, ensuring their matching with real-world entities. Achieving this correspondence enables a closed-loop system in which decisions concerning the physical entity are informed by real-time data derived from the DT.

In addressing the demands of modeling and designing radio maps tailored to the dynamic and complex environments in the forthcoming sixth-generation (6G) networks, the DT presents a compelling avenue. In fact, by leveraging high-definition three-dimensional (3D) maps, DT-enabled network systems can achieve precise real-time digital representations of the EM environment and radio maps. DT effectiveness can be improved by incorporating multimodal sensory data from different entities in the network, including user equipment, connected vehicles, and drones, in conjunction with a comprehensive description of the state of the communication network ~\cite{10198573}.

Wireless propagation is notably featured by multipath propagation. Ray-based propagation simulation stands as a versatile modeling tool, offering estimates of path loss, angle of arrival/departure, propagation delay, and Doppler shift for each multipath component. Functionally, ray-based propagation simulation relies on the high-frequency approximation of Maxwell's equations, which results in the concept of ray. Furthermore, it allows for effective integration with 3D scenarios maps to dynamically and faithfully model the propagation environment with rich geometric features \cite{7152831}. 

Different organizations have proposed computational engines to perform ray-based propagation simulations, and by integrating those software into the DT-enabled network systems, it is possible to better simulate the real-world conditions of the EM environment in urban settings, where the propagation of high-frequency signals is significantly affected by the urban landscape. By accounting for different possible interactions with the environment--- e.g., reflections, diffractions, and diffuse scattering--- we can achieve more accurate estimates and designs for wireless communication systems. Specifically in dynamic scenarios, such as vehicular ones, traditional models may fail to accurately represent channel characteristics \cite{9770941}. 
As research on DT-enabled network systems continues, it becomes incrementally more relevant to assess and understand their integration with ray-based propagation simulations.

Ray-based propagation simulations are known to be computationally expensive and therefore substantial research efforts have been directed toward enhancing computational speed \cite{9459462, 10089404, 8529268, BostonTWIN, 10238157}. In \cite{9459462} a simplification of simulations based on ray-based models is proposed and evaluated in end-to-end networks. An alternative dynamic ray-based method that is in \cite{10089404} alleviates the computational load of ray-based models. An adaptive ray launching algorithm is proposed for urban environments \cite{8529268}, reducing its computational burden. The high-fidelity 3D model of the city of Boston, introduced in \cite{BostonTWIN}, offers easier integration into DT-enabled systems. In \cite{10238157} a simplification of building representation is proposed for ray models.

The computational burden associated with ray-based propagation simulations can easily become the bottleneck towards achieving real-time performance in the update of a DT. In addition, the dynamism of the environment may render the simulation outdated even before the end of its computation.

%\textbf{Time Complexity.}
%Traditionally, the computational time of an algorithm is analyzed using time complexity analysis. This method involves examining the algorithm's behavior as a function of the input size. The time complexity is estimated by counting the number of elementary operations performed by the algorithm, assuming that each elementary operation takes a constant time. This function is generally hard to compute and requires access to the source code, which is not available for a proprietary software solution.
%The running time of the algorithm may vary among different inputs of the same size, so it is common to consider the worst-case time complexity. The focus is typically on the behavior of the complexity as the input size increases --- specifically, the asymptotic behavior of the complexity. Therefore, the time complexity is commonly expressed using the big O notation. This function is generally hard to compute and requires access to the source code. 

%\textbf{Computational Benchmark.} Time complexity analysis is nontrivial for algorithms composed by multiple dependencies and lines of code. 
% typically using Big O notation to classify its efficiency and scalability.

\textbf{Computational Benchmark.} Traditionally, the computational time of an algorithm is analyzed using time complexity analysis. This method involves examining the algorithm's behavior as a function of the input size. The time complexity is estimated by counting the number of elementary operations performed by the algorithm, assuming that each elementary operation takes a constant time. The running time of the algorithm may vary among different inputs of the same size, so it is common to consider the worst-case time complexity. 
Evaluating this function is typically challenging for complex software and requires access to the source code, which is not available for proprietary solutions.
Under these conditions, the next best option is to perform time measurements.
Thus, a benchmark for ray-based models has been proposed, which is the main contribution of our paper. Hardware, scenario, and inputs are fixed to measure the time performance of software implementations.

\textbf{Contributions.} 
In this paper, our aim is to provide a technical benchmark to evaluate the \textit{computational performance of different existing} ray-based simulation software. The main contributions are summarized below:
\begin{itemize}
    \item Definition of an empirical benchmark to evaluate and compare the computational performance of ray-based simulation software in two different urban scenarios characterized by their geometrical representation complexity.
    \item Evaluation of time performance for a selection of ray-based simulation software in the proposed urban scenarios and wireless link configurations--- i.e., single link vs. simultaneous multiple links; the latter is evaluated with GPU acceleration (when applicable). The evaluation has been carried out for increasing reflection bounces, with/without diffuse scattering, and variation of number of rays sampled at transmitter.
    \item The 3D meshes of the evaluation scenarios and the source code to replicate the benchmark for the evaluated solutions are available\footnote{Source code and scenarios are available in https://github.com/Michele-Zhu/ray-launching-benchmark.}. Fostering a repeatable benchmark with a standardized approach to evaluate the computational performance of diverse ray-based propagation algorithms, spanning different implementations and hardware architectures. 
\end{itemize}
\bigskip
By establishing shared simulation settings and scenarios, researchers and developers can have a comprehensive framework for in-depth comparison and analysis of the efficiency of their methodologies. The definition of a standardized methodology promotes transparency and repeatability in research, driving advances in ray-based propagation modeling, and addressing the evolving needs of wireless communication systems.

\bigskip
\textbf{Organization.} The remainder of this article is structured as follows: in Sec. \ref{sec:basics_of_ray_tracing}, we provide an overview of the basic ray-based simulation algorithms. Section \ref{sec: Benchmark framework} describes the 3D maps of the urban environments considered in our simulations, the type of wireless link configurations, and the proposed benchmark framework. Sec. \ref{sec:compared_ray_tracing_solutions} provides a brief review of software that implements ray-based models. In Sec. \ref{section: Numerical Simulation}, we report the simulation results on the comparison of the selected ray-based simulation software. Section \ref{sec: discussions} provides an overview and some remarks about the compared solutions, while in Sec. \ref{sec:conclusion}, we draw the conclusions. 

\section{Ray-based propagation algorithms} \label{sec:basics_of_ray_tracing} In this section, we differentiate the two main approaches to ray-based models and briefly describe the main implementations.

\subsection{Modeling approaches}
Following the distinction made in ~\cite{gschwendtner1995ray}, ray-based models can be differentiated into exact ray computation (called ray tracing or deterministic approach) and approximate (called ray launching or empirical approach):
\begin{itemize} [wide]
	\item \textbf{Ray Launching (RL)}, is an approximate model based on spawning rays from an EM emission source towards a set of angular discretized directions. For each launched ray, after several bounces (usually a configuration parameter) with the propagation environment, the reception of the ray at a target receiver is checked---e.g., using ray-tube modeling \cite{iskander2002propagation} or testing if the ray hits a sphere of given (possibly parametric) radius centered at the receiver position. The goal of \textit{plain} ray launching is not exact point-to-point propagation modeling. Instead, ray launching is particularly suitable for large-scale approximate propagation simulation. One realization of ray launching is the \textit{shooting and bouncing rays} method discussed below.
	\item \textbf{Ray Tracing (RT)},  aims at modeling point-to-point propagation by determining the geometrically exact propagation paths between an EM emission source and a target receiver location. According to Reif \textit{et al.} in \cite{10.1007/BF02574009}, \textit{``the exact solution of a 3D optical system consisting of linear reflective surfaces and partially reflective surfaces is undecidable"}. The computational time of the ray tracing task can easily become intractable with the increase in the number of surfaces to be taken into account or with the maximum number of interaction points per ray to be considered. The image method, discussed below, is a realization of the ray tracing method for reflection interaction types. 
\end{itemize}

%\medskip

\subsection{Ray-based Models Implementations}
Ray launching and ray tracing lead to a set of fundamental algorithms in computational EM modeling, among which are the main ones discussed below.

\begin{figure}
    \centering
    \subfloat[Image method]{\includegraphics[width = 0.5\columnwidth]{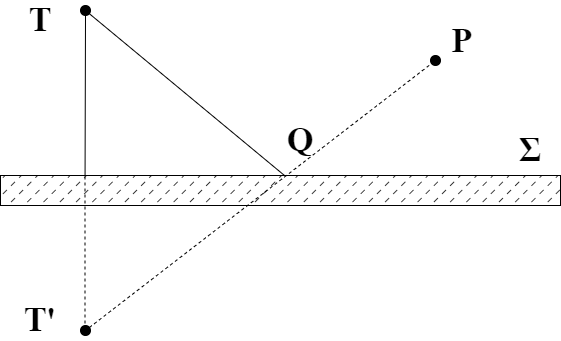}\label{fig: Image method}}
    \subfloat[Shooting and Bouncing Rays]{\includegraphics[width = 0.5\columnwidth]{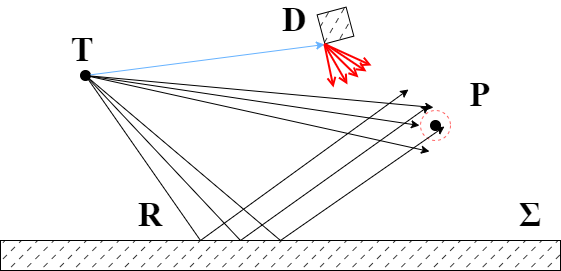}\label{fig: SBR}}
    \label{fig: RT methods}
    \caption{\protect\subref{fig: Image method} Image Method; T$^\prime$ is the image of transmitter T with respect to the reflection plane $\Sigma$. \protect\subref{fig: SBR} Shooting and Bouncing Rays method; R indicates reflection points, D a diffraction point at an edge, and a path is considered at the receiver if it satisfies the reception condition---exemplified by hitting a reception sphere (red circle) at receiver P.}
\end{figure}

\textbf{Image Method [RT]:} The image method is a point-to-point propagation simulation method. As such, it is a realization of the above-discussed ray tracing method. We provide in Fig. \ref{fig: Image method} a graphical representation of the image method for the single reflection case. 
Referring to Fig. \ref{fig: Image method}, the trajectory of the ray reflected from a plane surface can be determined as follows:
T$^\prime$ is computed as the image of transmitter T with respect to the reflection plane $\Sigma$; then T$^\prime$ and P are connected through a segment that intersects the plane at point Q; the reflected ray is determined by the segments connecting (T, Q, P). In case of multiple reflections, the method can be extended recursively taking into consideration different reflection planes.
The image method becomes computationally taxing when the environment presents an abundant number of reflecting surfaces or the number of sequential interactions considered for a ray increases \cite{7152831}. Nowadays, most RT tools implement Geometry Theory of Diffraction or Uniform Theory of Diffraction (GTD/UTD) \cite{92RS01781} to account for diffraction interactions of traced rays and in combination with reflections. Moreover, diffuse scattering can also be included in combination with other interactions, a RT tool with only reflection is not reliable. %\textcolor{red}{Image method alone is incapable of modeling interactions other than reflections, and the authors in \cite{92RS01781} show an RT implementation that also includes diffracted rays.}

\textbf{Shooting and Bouncing Rays (SBR) [RL]:} is a realization of the ray launching approach. A graphical representation is provided in Fig. \ref{fig: SBR}. A set of rays is launched from the transmitter T with a given angular separation. In conventional implementations, the launched rays are propagated in the environment up to a maximum number of interactions with surfaces and are considered to reach the receiver P if they meet a reception condition--- e.g., intersecting a sphere centered at the receiver P with a radius proportional to the distance traveled by the ray. SBR implementations can consider a variety of environmental interactions, e.g., reflection, diffraction, and diffuse scattering. The latter has been shown to be particularly relevant to accurately model urban propagation environments\cite{degli2001diffuse}.

\textbf{Path correction methods [RL and RT]:} In SBR, the \textit{exact} reception of a ray at a given target point in space is highly unlikely owing to the discrete angular resolution of ray launching. For this reason, path correction methods~\cite{7152831} have been proposed to exploit both the computational efficiency of SBR and the exact geometric accuracy of the image method. This is achieved by slightly changing the positions of the intermediate path interaction points on the surfaces so that the launched ray can \textit{exactly} hit the target point. Examples of path correction methods are implemented in the SBR models proposed by \cite{Wireless-InSite} and \cite{Matlab-raytracer}. For large propagation scenarios the overhead caused by the addition of path correction techniques to SBR is negligible \cite{7152831}. 

\textbf{Differentiable Ray Tracing [RL or RT]:} A new approach to ray modeling based on differentiable rendering (DR)~\cite{kato2020differentiable} has recently been proposed in \cite{hoydis2023sionna}. The developed procedure is based on Mitsuba 3 \cite{mitsuba} DR system--- which, in turn, leverages the Dr.Jit just-in-time (JIT) compiler \cite{Jakob2020DrJit}. DR for ray simulation brings as a key advantage differentiability with respect to environmental (e.g., objects' radio-materials parameters) and system (e.g., antenna arrays positions) of the ray tracing procedure. In \cite{hoydis2023sionna}, both an exhaustive ray tracing method and an approximate SBR-based approach are proposed. The exhaustive method tests all the possible combinations of mesh triangles and edges--- leading to high accuracy but intractable computational time. The ray launching method provides a trade-off between accuracy and complexity by means of a sampling scheme on the ray launching sphere. 
In \cite{eertmans2024fully}, the authors tackle the challenges of integrating ray tracing into DR systems, proposing a new fully differentiable framework with continuous loss function through local smoothing.

\textbf{Ray Tracing vs Ray Launching:} RT aims to model geometrical exact propagation paths, the image method checks each surface in the scene in order to find the propagation paths. RL, on the other hand, is an approximate model in which a finite number of rays are launched in discretized directions. RL inherently cannot find all the propagation paths. RT is preferred when high-fidelity and detailed analysis is required, such as point-to-point wireless channel modeling. RL is more suitable for coverage prediction and large-scale simulation in wireless communication.

\noindent According to \cite{529137} and \cite{9362202}, when considering only reflections up to $K$ bounces and non-optimized implementations, the image method has a computational complexity of $O(N^K)$, where $N$ is the number of planar surfaces. SBR has a computational complexity of $O(RK)$, where $R$ is the number of rays spawned. In a typical large propagation scenario, we have that $ N^K >> RK$. RT reflection bounces implemented with the image method can easily become intractable when the number of planar surfaces increases, thus making RT unsuitable for real-time applications and DT implementations.

%\noindent In ray-based model the most expensive computation is the ray search operation. Considering only reflection up to k-th bounces, in [] it has been demonstrated that with image method the number of propagation paths $M$ that have to be examined is exponential w.r.t. the number of reflecting surfaces $N$, $M_{RT} = 1 + \sum_{i=1}^{k} N(N-1)^{i-1}$. While RL with ray-tube modeling has to examine $M_{RL} = \lceil 4 \pi \cdot (k + 1)^2 \cdot (\frac{d_{max}}{\Delta})^2 \rceil$, where $d_{max}$ is the maximum length of a straight line that lies inside the considered geometry, 

%When considering only reflections up to k-th order bounces it has been shown that the number of paths for image method is exponential w.r.t. the number N surfaces reflecting surfaces. While SBR with ray tube model d

%\noindent RT typically requires more computational resources because of the complexity of tracing multiple paths and interactions. RL can be more efficient, especially with optimizations and approximations. 
    \color{black}

\vspace{-0.3cm}\section{Benchmark framework} %\ref{sec: benchmark framework}
\label{sec: Benchmark framework}
\begin{figure*} [!t]
    \centering
    \subfloat[Scenario1]{\frame{\includegraphics[width =0.45\textwidth]{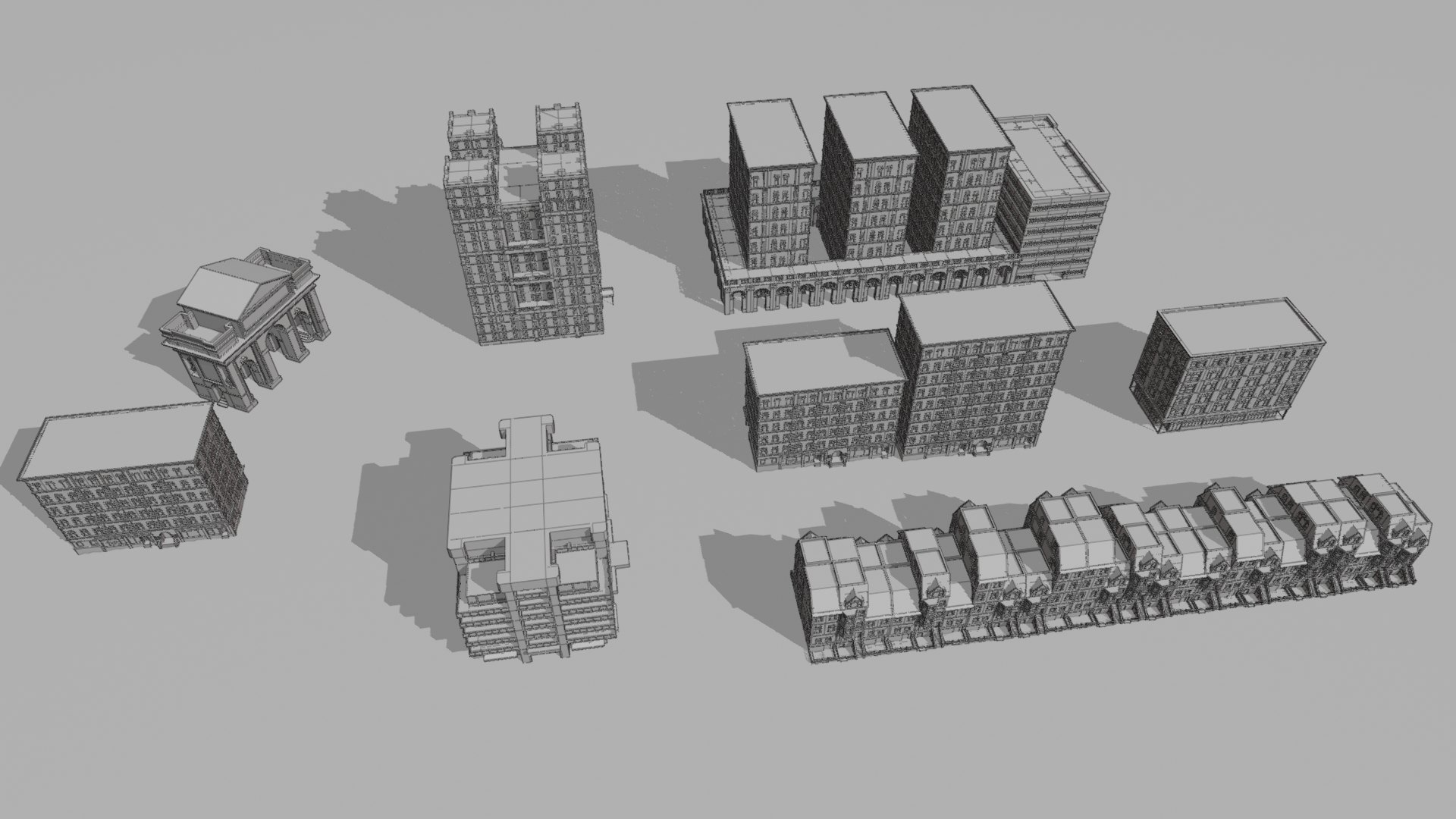}\label{fig: Wireframe Scenario1}}}
    \hspace{0.5cm}
    \subfloat[Scenario2]{\frame{\includegraphics[width = 0.45\textwidth]{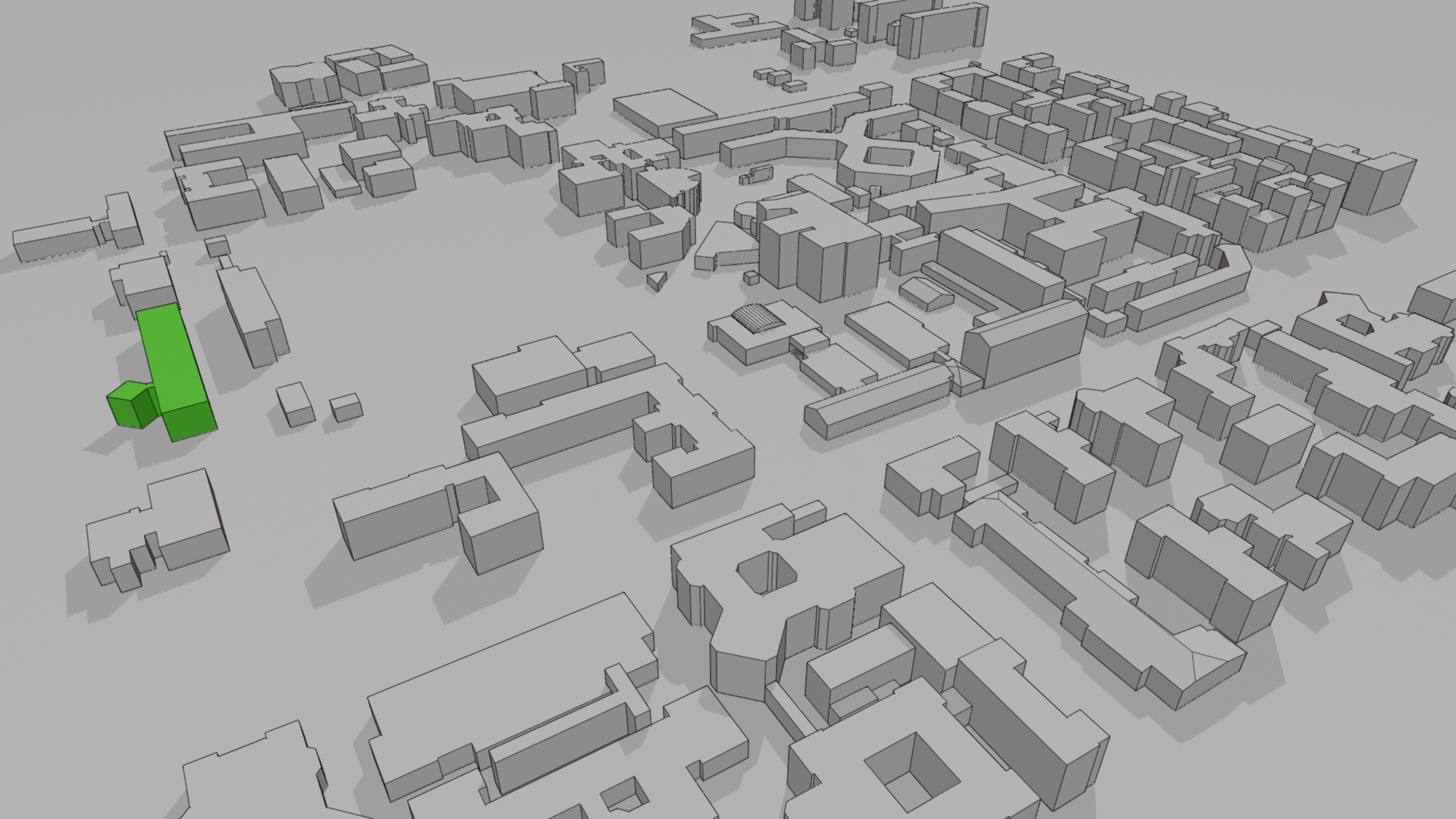}\label{fig: Wireframe Scenario2}}}
    \caption{Details of the 3D meshes of the two proposed scenarios. The view in \protect\subref{fig: Wireframe Scenario1} highlights the level of detail in Carla w.r.t. \protect\subref{fig: Wireframe Scenario2}, which is the view of the OSM scenario characterized by a lower level of details. In \protect\subref{fig: Wireframe Scenario2}, the main building of the High Frequency campus, Italy is in green.}
    \label{fig:Wireframe-view}
\end{figure*}
\begin{figure}[!t]
    \centering
    \subfloat[Scenario1]{\includegraphics[width = 0.9\columnwidth]{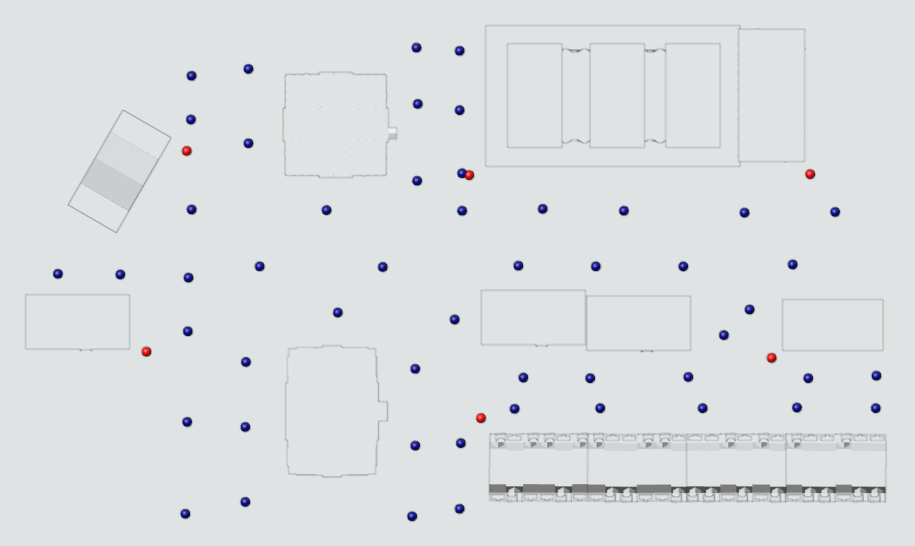}}
    
    \subfloat[Scenario2]{\includegraphics[width = 0.9\columnwidth]{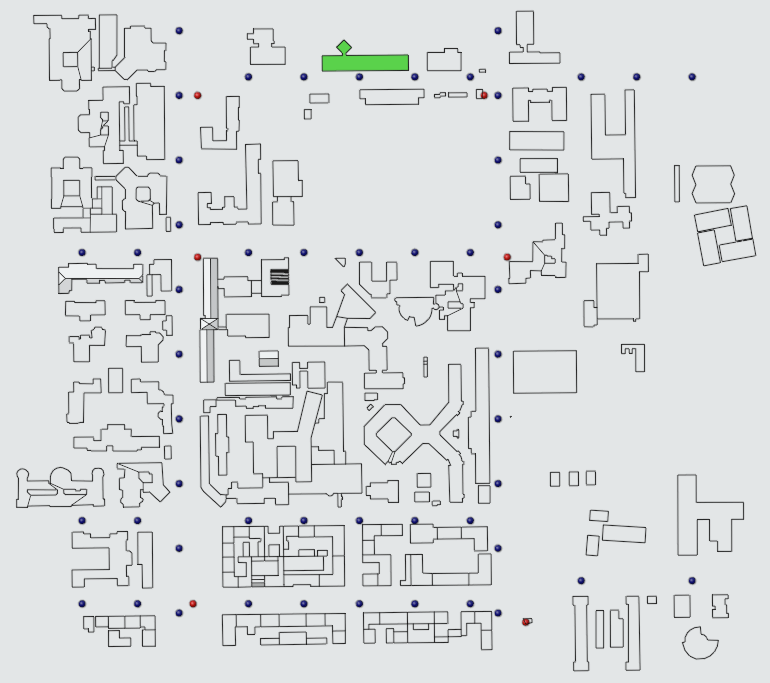}}
    \caption{Top views of the selected evaluation scenarios along with the chosen Tx (red spheres) and Rx (blue spheres) positions. In (b), the main building of the High Frequency Campus in Milan, Italy is in green.}
    \label{fig: top-view at vehicular position}
\end{figure}

In what follows, we describe urban scenarios, wireless link configurations, and the proposed benchmark framework to evaluate the ray-based method considered.

\subsection{Evaluation scenarios}
We have selected two urban evaluation scenarios, differentiated by the level of complexity of the buildings' meshes by which they are composed. Figure \ref{fig:Wireframe-view} shows the mesh complexity of the proposed scenarios.
\begin{itemize} [wide]
	\item \textbf{CARLA High Definition Mesh (Scenario1):} It is an environment composed by complex realistic 3D buildings' meshes retrieved from the \textit{Town 10} urban scenario available within the realistic CARLA automotive simulator \cite{Carla}. Town 10 represents a typical inner-city environment, composed of, e.g., skyscrapers, hotels, public buildings, and apartment blocks. We have considered a selection of buildings in Town 10 that led to a diversified urban environment, and we exported the corresponding 3D meshes using the CARLA Unreal Engine editing interface. The resulting 3D mesh is represented in Fig. \ref{fig: Wireframe Scenario1}. Average building height is 27.21 m.
	\item \textbf{OSM Low Definition Mesh (Scenario2):} As the second test environment, we consider a scenario composed by simplified 2.5D buildings' meshes---i.e., meshes defined by their outline on the ground and their height---retrieved from the OpenStreetMap (OSM) \cite{openstreetmap} online service for the High Frequency Campus urban geographical area in Milan, Italy. A section of the scenario is reported in Fig. \ref{fig: Wireframe Scenario2}, showing the simplicity of the building representation in this setting. Average building height is 14.29 m.
\end{itemize}

\subsection{Wireless link configurations}\label{sec: wireless_link_sim}
Let $\mathcal{G}$ be a generic set of transceiver configurations, $\mathcal{T}$ be the set of transmitters, $\mathcal{R}$ be the set of receivers. 
A generic transceiver configuration is defined as $g_k \coloneq a_k \cup b_k$ where $a_k \subseteq \mathcal{T}$ and $b_k \subseteq \mathcal{R}$.
Denoting $N_{Tx} = |\mathcal{T}|$, $N_{Rx} = |\mathcal{R}|$ the number of transmitters and receivers.

We define two distinct link configurations, represented by $\mathcal{G}_1$ and $\mathcal{G}_2$:

\begin{itemize} [wide]
    \item \textbf{Single Link (SL) configurations:} the set $\mathcal{G}_1$ represents the set of all Tx/Rx pairs. Ray computations and wall-clock time measurement are performed for $g_{ij}$ with $i = 1, 2, ..., N_{Tx}$ and $j = 1, 2, ..., N_{Rx}$. This approach captures the performance of the ray model without parallelization. 

    \item \textbf{Multi Link (ML) configurations:} a transceiver configuration $g_i \in \mathcal{G}_2$ is defined as $g_{i} = T_i \cup \mathcal{R}$ with $i = 1, 2, ..., N_{Tx}$. Measurements and computations are performed for the configurations in $\mathcal{G}_2$ simulating a channel propagation scenario with one transmitter and multiple receivers. This allows the different ray methods to exploit hardware parallelization and algorithmic optimization in realistic use cases.
\end{itemize}
\bigskip

Generic link configurations can be defined and characterized by their set of transceiver configurations $\mathcal{G}$, each different configuration can be designed to suit different use cases and capture different properties of the ray methods.
The algorithm in \ref{alg: simulation pseudocode} is the pseudocode for a generic simulation with set $\mathcal{G}$, measurements of the wall-clock time are performed after the transceivers in $g_k \in \mathcal{G}$ are loaded into memory. If round-trip-time measurements are more desirable, include the loading of the configuration $g_k$ within the simulation software.

The $N_{Tx} = 6$ transmitters are placed in nodal positions in urban environments at a height of 7 m, while the $N_{Rx} = 51$ receivers are distributed along roads at a height of 1.5 m. Each transmitter (Tx) and each receiver (Rx) are equipped with the same type of isotropic antennas, propagation mainly occurs on the horizontal plane w.r.t. the buildings. In Fig. \ref{fig: top-view at vehicular position} the positions of the transmitters (red spheres) and receivers (blue spheres) are shown.
%is shown the graphical depiction of transmitters (red spheres) and receivers (blue spheres).
When required, the simulation boundary was set with a 50 m margin with respect to the bounding box comprising all buildings. The simulation boundary is assumed to be absorbing, so that only rays interacting with the buildings and ground within the scenario are considered. The ITU recommendation \cite{ITU-R-2040} has been considered as reference to determine the required EM properties (relative permittivity and conductivity) of the radio materials. The simulation parameters for each scenario are summarized in Table \ref{tab:sim_param}.
%We summarize the general simulation parameters, common to all the simulation software considered, in Table \ref{tab:sim_param}.
\begin{algorithm}[!htb]
\begin{algorithmic}[1]
\Require $\mathcal{T}\coloneq$ set of transmitters, $\mathcal{R}\coloneq$ set of receivers.\newline $\mathcal{G} \coloneq$ set of transceiver configurations, $g_k = a_k \cup b_k$ with $g_k \in \mathcal{G}$, $a_k \subseteq \mathcal{T}$, and $b_k \subseteq \mathcal{R}$.
\Ensure $\boldsymbol{t}$ vector of wall-clock time measurements.
\State Initialize $\boldsymbol{t} \in \mathbb{R}^{|\mathcal{G}|}$
\ForAll{$g_k \in \mathcal{G}$}
    \State load\_transmitters($a_k$)
    \State load\_receivers($b_k$)
    \State $t^\prime=$ current\_timestamp()  
    \State compute\_paths($g_k$)
    \State $t_k=$ current\_timestamp() $-\ t^\prime$     
\EndFor
\State \Return $\boldsymbol{t}$
\end{algorithmic}
\caption{Simulation pseudocode for generic set $\mathcal{G}$}.
\label{alg: simulation pseudocode}
\end{algorithm}
\vspace{-0.5cm} % reduce default spacing between alg 1 and table 1

\begin{table} [!hbt]
    \centering
    \footnotesize
    \caption{Simulation parameters.} 
    \begin{tabular}{l c}
		\toprule
		\textbf{Parameter} & \textbf{Value} \\ %[2 pt]
		\noalign{\smallskip}
		\hline
		\noalign{\smallskip}
		Antenna Type & Isotropic \\\noalign{\smallskip}
		Carrier frequency & 28 GHz \\\noalign{\smallskip}
		Buildings and ground material & Concrete\\\noalign{\smallskip}
		Material relative permittivity & 5.31 \\\noalign{\smallskip}
		Material conductivity & 0.4838 S/m \\\noalign{\smallskip}
		Tx height, Rx height & 7 m , 1.5 m\\\noalign{\smallskip}
        $N_{Tx}$, $N_{Rx}$ & 6, 51 \\ \noalign{\smallskip}
		%Rxs height & 1.5 m\\\noalign{\smallskip}
        %$N_{Rx}$ & 51 \\\noalign{\smallskip}
        Scenario1 avg. building height & 27.21 m \\ \noalign{\smallskip}
        Scenario2 avg. building height & 14.29 m \\ \noalign{\smallskip}
		\bottomrule
    \end{tabular}
    \label{tab:sim_param}
\end{table}

\subsection{Benchmark workflow}
\begin{figure}
    \centering
    \includegraphics[width=0.9\linewidth]{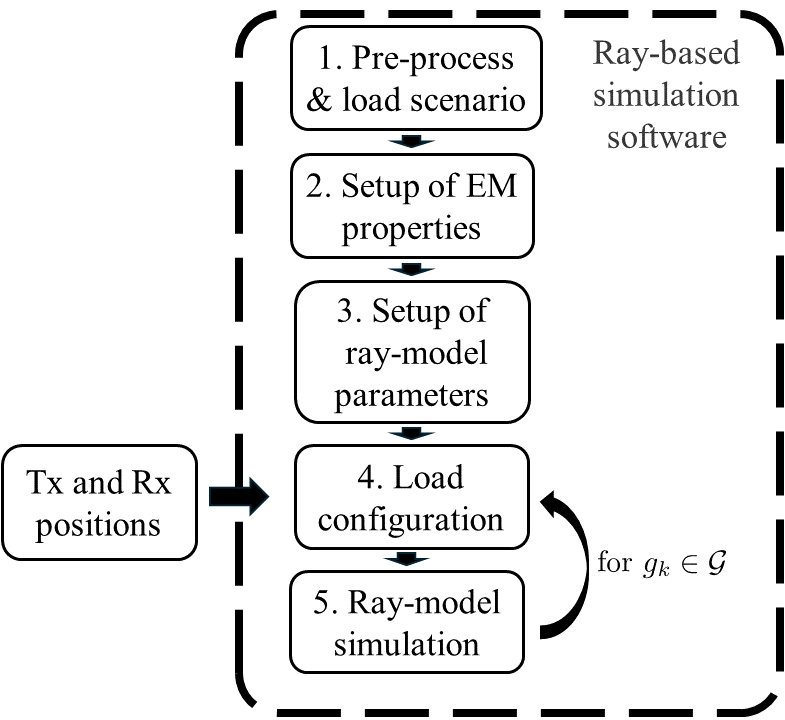}
    \caption{Benchmark workflow. $\mathcal{G}$ is a set of transceiver configurations and $g_k$ is a specific configuration containing transmitters and receivers.}
    \label{fig: Benchmark workflow}
\end{figure}
Given a ray-based simulation software, a scenario, and a configuration $\mathcal{G}$. The benchmark workflow can be described as follows.

\textbf{1. Pre-process \& load scenario.} Ray-based model simulation requires 3D meshes of the environment as input. These meshes may not be in a format compatible with the specific ray-based simulation software. If necessary, the scenario is converted from its original format into a supported one before loading it within the simulation software. This ensures that the simulation software can accurately interpret and utilize the environmental data.

\textbf{2. Setup of EM properties.} In this step, the EM properties of the scenario are configured. This includes defining parameters such as permittivity, permeability, and conductivity of materials within the simulation environment, according to the specifications in Table \ref{tab:sim_param}.

\textbf{3. Setup of ray model parameters.} Different software implementations of ray models may offer various configurable parameters. This step involves aligning and configuring these parameters to ensure consistency across different simulations.

\textbf{4. Load configuration.} Transmitters and receivers in the configuration are loaded within the simulated scenario at positions provided by an external file. 

\textbf{5. Ray-model simulation.} This step is the core computational step where the actual simulation of the ray model is performed. The software computes the paths that rays take as they propagate from the transmitters to the receivers. The wall-clock time is measured in this step\footnote{In order to capture steady-state performance a warm-up procedure is required. This means running the simulation for a few minutes without performing time measurements.}.

Steps 4. and 5. are repeated until the configurations $g_k \in \mathcal{G}$ are exhausted. Figure \ref{fig: Benchmark workflow} summarizes the benchmark workflow.

\section{Comparison of ray-based simulation software}\label{sec:compared_ray_tracing_solutions}

In this section, we review the examined ray-based simulation software. Open-source and commercial software embeds ray-based models for EM propagation modeling. Among the most widely used commercial software are Remcom Wireless InSite \cite{Wireless-InSite}, MathWorks RF Antenna Toolbox \cite{Matlab-raytracer}, Siradel Volcano \cite{Siradel-Volcano}, Altair Feko \cite{Altair-Feko}, iBware Design and iBware Reach \cite{iBwave}, and EDX SignalPro \cite{EDX-SignalPro}. NVIDIA Sionna \cite{sionna} and Ns-3 mmWave Module \cite{Ns-3mmWave} instead provide open-source solutions. 
Based on the availability of proprietary software, we evaluate the following ray-based simulation software.

\textbf{Remcom Wireless InSite (v3.3.3)} \cite{Wireless-InSite} is a commercial software that provides several methods for the analysis of radio propagation and wireless communication systems in various conditions (i.e., indoor, outdoor, and rural areas). Widely used for wireless propagation simulation in the relevant literature \cite{zaal2020optimal,jacovic2022experimentation}, it provides efficient and accurate modeling of the characteristics of the communication channel in complex EM propagation environments. It offers two ray optical computation models allowing generic 3D environments: (i) full 3D, supporting simulations in the 0.1-20 GHz frequency range by means of SBR and Eigen Ray algorithms, and (ii) X3D, which supports simulations in the 0.1-100 GHz frequency range and integrates SBR with a path correction algorithm adjusting the interaction points to obtain exact propagation paths between Tx and Rx. Moreover, X3D supports diffuse scattering simulation, which has been shown to be particularly relevant for accurately modeling urban propagation environments \cite{degli2001diffuse}.

\textbf{NVIDIA Sionna RT (v0.16.2)} \cite{hoydis2023sionna} is a recently proposed differentiable ray optical engine that is part of the NVIDIA Sionna link-level simulation library \cite{sionna}. NVIDIA Sionna RT allows for accurate ray simulations taking account of reflection, diffraction, and diffuse scattering interactions of propagation paths with the environment. As reported in Sec. \ref{sec:basics_of_ray_tracing}, it implements differentiable rendering providing a flexible framework that allows the inclusion of the communication channel within end-to-end optimization procedures. Sionna RT provides two ray-based methods: (i) an exhaustive method, which tests all possible combinations of 3D primitives and paths and becomes computationally intractable for high path depths---i.e., number of interaction points per path---or number of surfaces in the scene, and (ii) a \textit{Fibonacci} method, which uses the SBR approach to efficiently compute the propagation paths and implements a sampling procedure for ray launching that enables a trade-off between simulation accuracy and computational time.

\textbf{MathWorks Ray tracing model (vR2023a U1)} \cite{Matlab-raytracer} is part of the MathWorks Antenna Toolbox and offers two ray optical computation methods: (i) SBR implementation with exact path correction, supporting up to 10 reflection and 2 edge diffractions and providing an approximate number of propagation rays; (ii) a ray tracing model based on the image method, supporting a maximum of 2 path reflections and providing an exact number of rays featured by line-of-sight or reflection with exact geometry. Both methods enable simulation in the 0.1-100 GHz frequency range and support generic 3D indoor and outdoor environments.

\section{Numerical results} \label{section: Numerical Simulation}

In this section, we evaluate the computational time efficiency of the considered ray-based methods in Scenario1 and Scenario2. %Moreover, the prediction of the radio map obtained by the different solutions has been also compared. 
In addition, we generate the radio map from different simulation software.
Our simulations are performed on a Windows 10 workstation equipped with Intel(R) Core(TM) i7-9700K CPU@3.60 GHZ, 8 cores, 16 GB RAM, and NVIDIA GeForce GTX 1070 Ti GPU with 8GB of dedicated memory.

Benchmark results are dependent on the underlying hardware infrastructure and how effectively an algorithm is optimized for it. In cases such as a dynamic Digital Twin deployment \cite{10198573} where time constraints are very important, this benchmark should be reproduced for the actual computing infrastructure and the input of the ray-model simulation by defining the appropriate set of Tx/Rx configurations $\mathcal{G}$ and 3D meshes representing the environment. When comparing the performance of different implementations of the ray-based model, the sets $\mathcal{G}_1, \mathcal{G}_2$ and our scenarios can be used instead.

%Ray simulation's computational performance hinges on algorithmic approach, ray pruning, hardware acceleration, and scenario complexity.
%We measure the wall-clock time, which is the system time elapsed during ray-based propagation simulation on a chosen system architecture.

\subsection{Computational Time}

\begin{figure*}[!t]
    \centering
    \subfloat[SL Scenario2: WI X3D]{
        \includegraphics[width=0.33\linewidth]{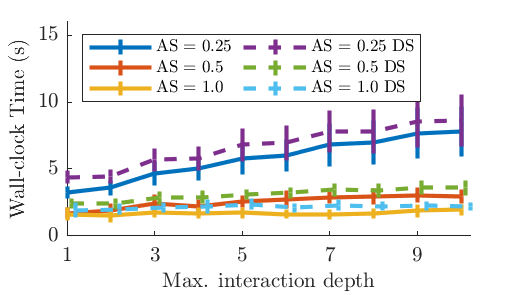}\label{subfloat: SL Scenario 2 WI}
    }
    \subfloat[SL Scenario2: Sionna Fibonacci CPU]{
        \includegraphics[width=0.33\linewidth]{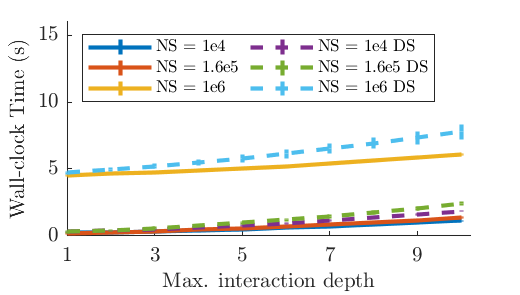}\label{subfloat: SL Scenario 2 Sionna}
    }    
    \subfloat[SL Scenario2: MW SBR]{\includegraphics[width=0.33\linewidth]{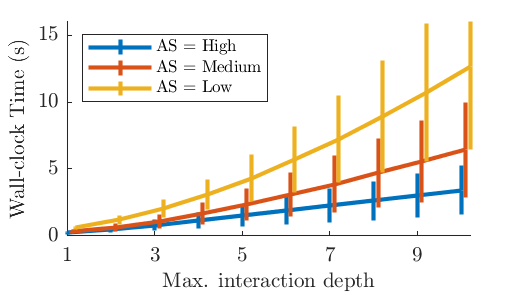}\label{subfloat: SL Senario 2 matlab}}  

    \subfloat[ML Scenario2: WI X3D]{\includegraphics[width=0.33\linewidth]{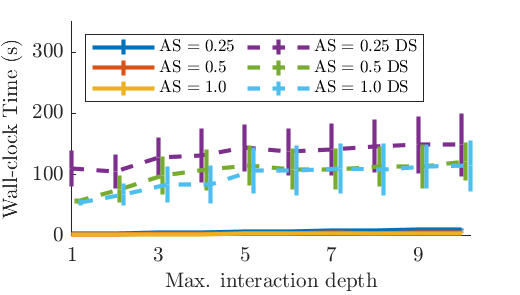}
        \label{subfloat: ML Scenario 2 WI}
    }
    \subfloat[ML Scenario2: Sionna Fibonacci GPU]{
        \includegraphics[width=0.33\linewidth]{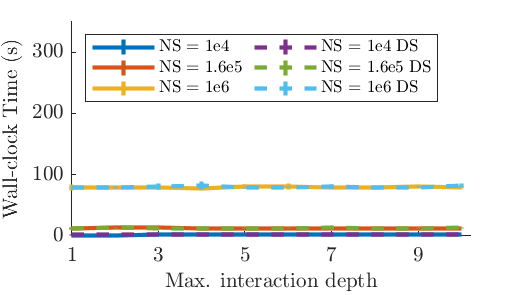}
        \label{subfloat: ML Scenario 2 Sionna}
    }
    \subfloat[ML Scenario2: MW SBR]{\includegraphics[width=0.33\linewidth]{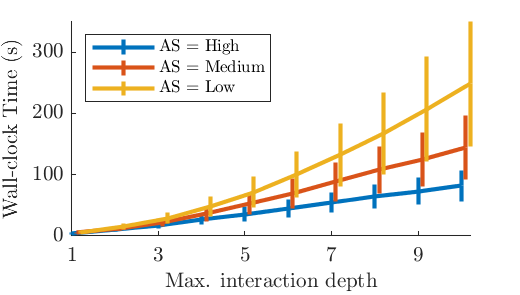}\label{subfloat: ML Senario 2 matlab}}

    \caption{\textit{Single Link (SL)} and \textit{Multi Link (ML)} configurations in Scenario2 with 1$\sigma$ errorbar with respect to the maximum interaction depth, varying \textit{angular separation} (AS) or \textit{number of samples} (NS), with (solid line) or without (dashed line) Diffuse Scattering (DS).}
    
    %\caption{\textit{Single Link (SL) configurations} in Scenario2 with 1$\sigma$ errorbar with respect to the maximum interaction depth, varying \textit{angular separation} (AS) or \textit{number of samples} (NS), with or without Diffuse Scattering (DS).}
    \label{fig: simulation_scenario2}
\end{figure*}

\begin{figure}[!t]
    %\vspace{-0.5cm}
    \centering 
    \subfloat[SL Scenario1: WI X3D]{
        \includegraphics[width=.75\columnwidth]{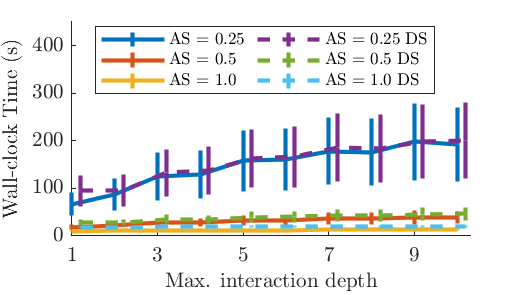}\label{subfloat: SL Scenario1 WI}
    } 
    
    \subfloat[SL Scenario1: Sionna Fibonacci CPU]{
        \includegraphics[width=.75\columnwidth]{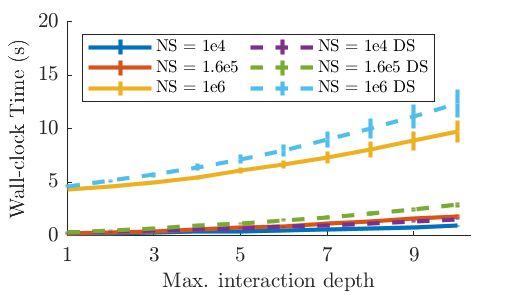}%{assets/single_link_simulations/sl_carla_sionna.png}%
    }   
    \caption{\textit{Single Link configurations} in Scenario1 with 1$\sigma$ errorbar with respect to the maximum interaction depth, varying \textit{angular separation} (AS) or \textit{number of samples} (NS), with (solid line) or without (dashed line) Diffuse Scattering (DS).}
    \label{fig: simulation-sequential-scenario1}
    %\vspace{-0.25cm}
\end{figure}
\begin{figure} [!t]
    %\vspace{-0.5cm}
    \centering
    \subfloat[Scenario1: WI X3D]{
        \includegraphics[width=0.75\columnwidth]{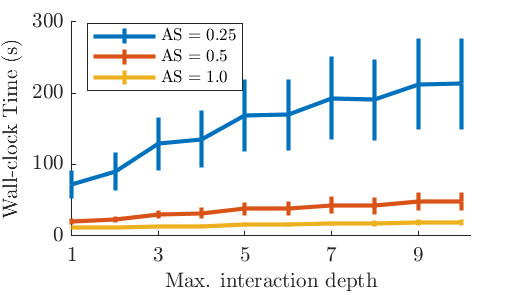}
        \label{fig: simulation-parallel-scenario1-a}
    }
    
    \subfloat[Scenario1: Sionna Fibonacci GPU]{
        \includegraphics[width=0.75\columnwidth]{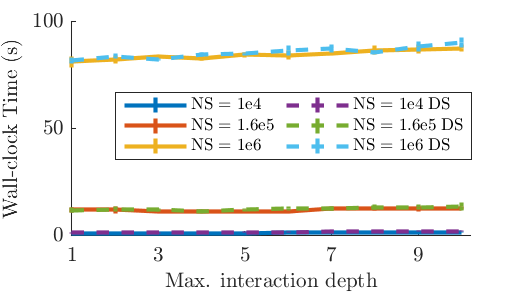}
        \label{fig: simulation-parallel-scenario1-b}
    }
        
    \caption{\textit{Multiple Link configurations} in Scenario1 with 1$\sigma$ errorbar with respect to the maximum interaction depth, varying \textit{angular separation} (AS) or \textit{number of samples} (NS), with (solid line) or without (dashed line) Diffuse Scattering (DS).}
    \label{fig: simulation-parallel-scenario1}
    %\vspace{-0.25cm}
\end{figure}

As discussed in Sec. \ref{sec: Benchmark framework} we consider a set of transmitters $\mathcal{T}$, a set of receivers $\mathcal{R}$. The wall-clock time $t_k$ is measured during the ray computation, for each transceiver configuration $g_k \in \mathcal{G}$. Sample mean $\mu$ and variance $\sigma^2$ are evaluated for the specific \textit{Single Link} (SL) and \textit{Multi Link} (ML) simulation, with $K = |\mathcal{G}|$ as:
\begin{equation}
    \begin{aligned}
        \mu &= \frac{1}{K} \sum_{k=1}^{K}  t_{k}, \\
        \sigma^2 &= \frac{1}{K - 1} \sum_{k=1}^{K} (t_{k} - \mu)^2.
    \end{aligned}
\end{equation}

\begin{figure*} [!t]
    \centering
    \subfloat[\textit{Single Link configurations}  on Scenario1]{
        \includegraphics[width = \columnwidth]{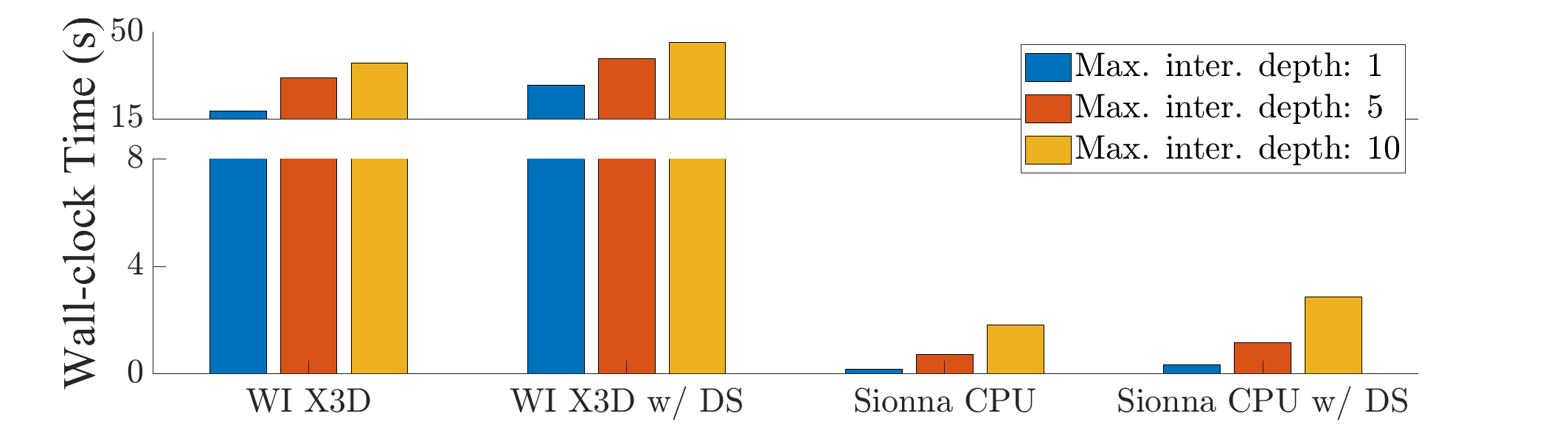}
        \label{fig: comparison-sequential-scenario1}
        }
    \subfloat[\textit{Single Link configurations} on Scenario2]{
        \includegraphics[width = \columnwidth]{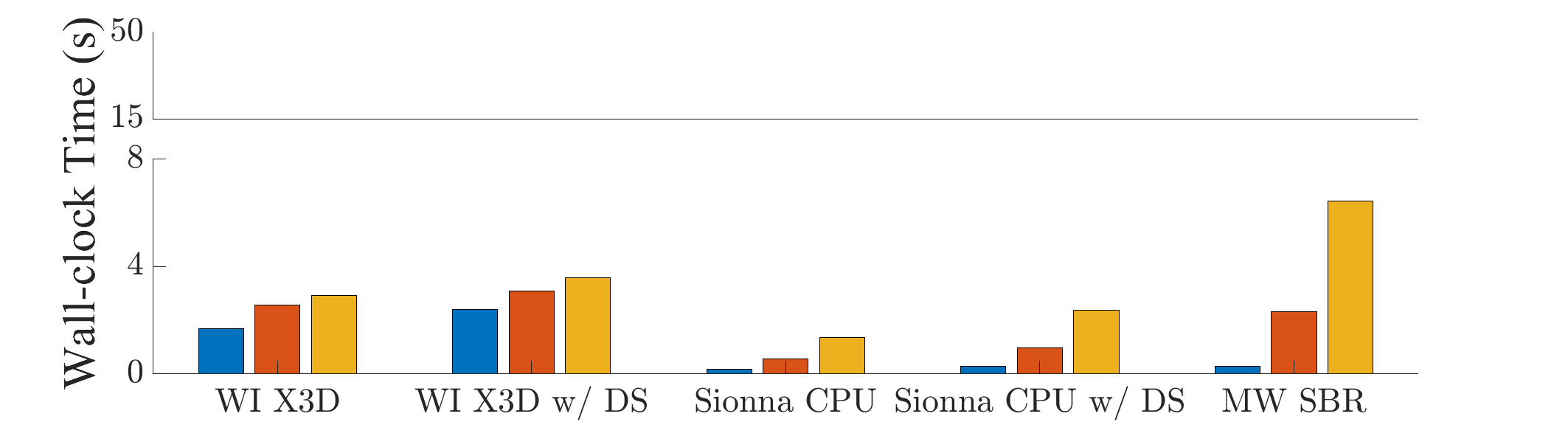}
        \label{fig: comparison-sequential-scenario2}}
    
    \subfloat[\textit{Multiple Link configurations}  on Scenario1]{
        \includegraphics[width = \columnwidth]{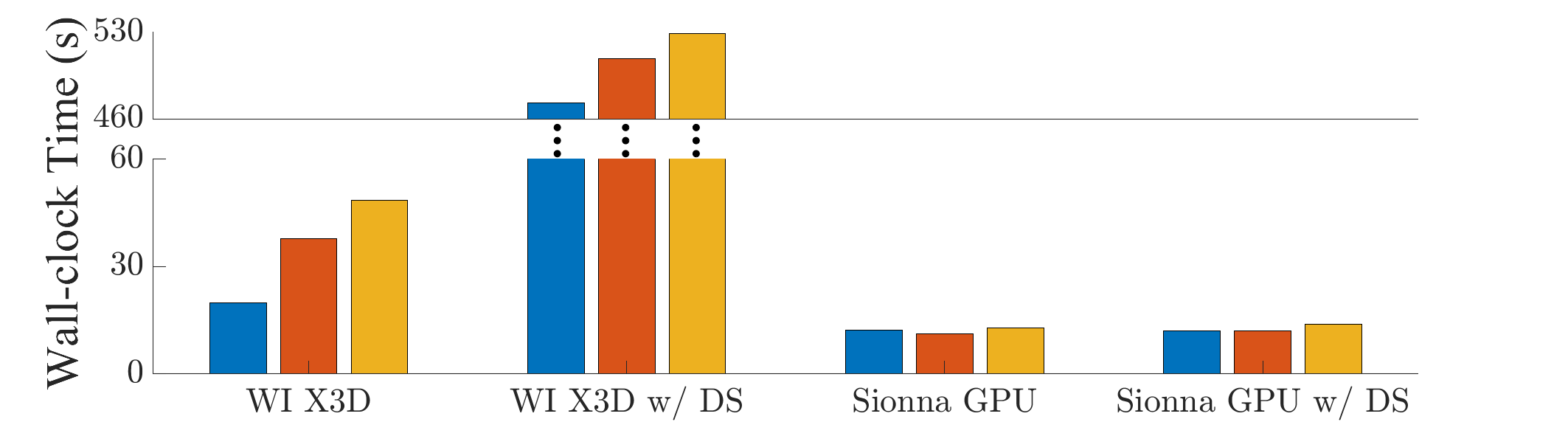}
        \label{fig: comparison-parallel-scenario1}}
    \subfloat[\textit{Multiple Link configurations}  on Scenario2]{
    \includegraphics[width = \columnwidth]{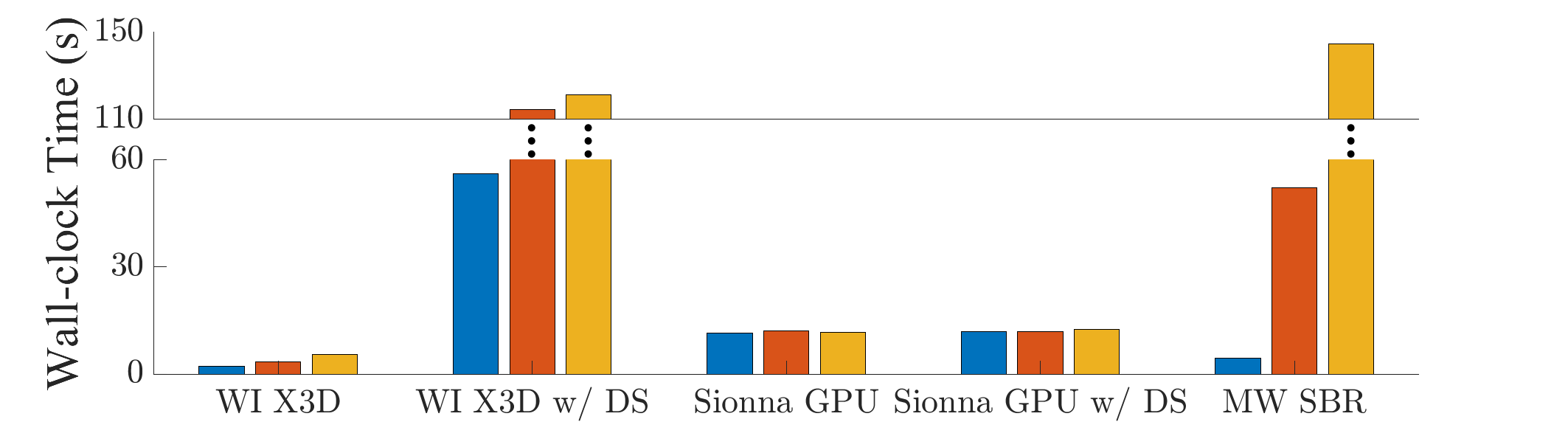}
    \label{fig: comparison-parallel-scenario2}}
    \caption{Comparison of the selected ray-based simulation software over a comparable initial number of rays at Tx. WI X3D AS $=0.5$ deg, MW SBR AS = Medium, Sionna Fibonacci NS $=1.6e5$.
    % \textit{ray launching parameter} (RLP) setting of 0.5 deg. 
   }
    \label{fig: Comparison simulations}
\end{figure*}

We consider the X3D model for \textit{Remcom Wireless InSite} (WI X3D), the SBR with exact path correction model for the \textit{MathWorks ray launching model} (MW SBR), and the \textit{Fibonacci} model for NVIDIA Sionna RT. MW SBR has been performed within its \textit{Graphical User Interface} (GUI), adding a negligible overhead in terms of computational time. MW SBR cannot handle Scenario1 on our workstation. 

CPUs and GPUs are different types of hardware resources. The former are optimized to handle sequential tasks and general-purpose computing. The latter are particularly useful for handling tasks that require parallel execution and high throughput--- Sionna framework allows for finer control of the type of resource used in the computation, in SL configurations we do not enable GPU acceleration, while in ML configurations we enable it. 

Ray launching algorithms are significantly impacted by the initial number of rays sampled at the Tx. Both WI X3D and MW SBR offer the flexibility to specify the \textit{angular separation} (AS) in degrees between the launched rays, while Sionna Fibonacci allows for the specification of the \textit{number of samples} (NS)--- e.g., the number of candidate rays split equally among the Tx during the simulation.
Another key parameter that influences the computational time is the type of interactions enabled and the number of bounces. We have to disambiguate the parameters in the Sionna Fibonacci model: the model provides Boolean parameters to enable/disable the type of interaction and to set the maximum interaction depth to set the maximum number of bounces jointly for all the enabled interaction types.
% and set the maximum interaction depth to set the maximum number of bounces for any type of interaction.
% WI X3D and MW SBR allow us to set the maximum number of bounces separately for each type of interaction. It is worth stating that when we use max. interaction depth for WI X3D and MW SBR we are referring to maximum number of reflection bounces.
% It is worth stating that in the definition max. interaction depth for WI X3D and MW SBR accounts for the maximum number of reflection bounces of a single ray.
WI X3D and MW SBR allow us to set the maximum number of bounces separately for each type of interaction. It is important to note that in the provided figures, the maximum interaction depth for WI X3D and MW SBR refers to the maximum number of reflection bounces of a single ray.
%(we abuse the name and perform simulations with increasing number of reflection bounces, but call it the maximum interaction depth in the provided figures).
Simulations are performed for the two proposed scenarios, SL/ML configurations, and with diffuse scattering enabled/disabled when applicable. The varying parameters for each of the ray launching methods are summarized in Table \ref{tab:rt_param}, while the unmentioned parameters are set to default values. 

\begin{table} [!bth]
	\centering
	\caption{Software-specific ray launching simulation parameters, reflection and diffraction bounces are always considered. Where possible we perform simulations with diffuse scattering enabled/disabled.}    
	\begin{tabular}{r l c}
		\toprule
		\textbf{Software} & \textbf{Parameter} & \textbf{Value}\\ %[2 pt]
		\noalign{\smallskip}
		\hline
		\noalign{\smallskip}
		\multirow{4}{*}{\shortstack[r]{Remcom\\ Wireless InSite\\ (v3.3.3)}} & RT method & X3D \\\noalign{\smallskip}
		& Angular separation & [0.25, 0.5, 1.0] deg\\\noalign{\smallskip}
            & Reflections bounces & from 1 up to 10\\ \noalign{\smallskip}
		& Diffraction bounces & 1 \\\noalign{\smallskip}
		& Diffuse scattering & $(R=2, D=1)$ \\\noalign{\smallskip}
		\midrule
		
		\multirow{4}{*}{\shortstack[r]{MathWorks\\ Ray Tracing\\ Model \\ (vR2023a U1)}} & RT method & SBR \\\noalign{\smallskip}
		& Angular separation & [low, medium, high]\\\noalign{\smallskip}
            & Reflections bounces & from 1 up to 10\\ \noalign{\smallskip}
		& Diffraction bounces & 1 \\\noalign{\smallskip}
            & Diffuse Scattering & N/A\\ \noalign{\smallskip}
		\midrule
		
		\multirow{4}{*}{\shortstack[r]{NVIDIA\\ Sionna RT \\ (v0.16.2)}} & RT method & Fibonacci \\\noalign{\smallskip}
		& Num. of samples & [1e4, 1.6e5, 1e6]\\\noalign{\smallskip}
            & Max. interaction depth & from 1 up to 10\\ \noalign{\smallskip}
            & Reflection flag & True\\ \noalign{\smallskip}
		& Diffraction flag & True \\\noalign{\smallskip}
		& Diffuse scattering flag & False/True \\\noalign{\smallskip}
		
		\bottomrule
	\end{tabular}
	
	\label{tab:rt_param}
\end{table}

%In Fig. \ref{fig: simulation_scenario2}, the results of the SL and ML configurations are reported for Scenario2. For SL configurations, it can be observed that for WI X3D and Sionna CPU the addition of diffuse scattering interaction adds a negligible overhead in the computation. MW SBR does not model diffuse scattering. For ML configurations, the WI X3D model observes a significant increase in computational time, Sionna GPU is also not influenced by the addition of diffuse scattering interaction.
In Fig. \ref{fig: simulation_scenario2}, the results of the SL and ML configurations are reported for Scenario2, MW SBR does not model diffuse scattering. For SL configurations, it can be observed that for WI X3D and Sionna CPU the addition of diffuse scattering interaction adds a negligible overhead in the computation. For ML configurations, the WI X3D model observes a significant increase in computational time, for Sionna GPU the overhead added by diffuse scattering cannot be visually observed.

In Fig. \ref{fig: simulation-sequential-scenario1}, the results of the SL configurations for Scenario1 are reported. Scenario1 is characterized by a higher number of meshes--- i.e. the number of vertices, edges, and faces that define the shape of the objects in the 3D map. This complexity significantly impairs the ray computations. An inspection between Fig. \ref{subfloat: SL Scenario 2 WI} and
\ref{subfloat: SL Scenario1 WI} highlights the impact of mesh complexity, specifically when diffuse scattering is considered.

Fig. \ref{fig: simulation-parallel-scenario1}, presents the results of ML configurations for Scenario 1. 
Fig. \ref{fig: simulation-parallel-scenario1-a} reports the computation for WI X3D without diffuse scattering. When diffuse scattering is considered the overall average wall-clock time: is 440 s with maximum interaction depth set to 1 and AS 1.0 deg, while 730 s with maximum interaction depth set to 10 and AS 0.25 deg. 

The initial number of rays significantly impacts the computational performance of the ray launching approach. In order to provide a direct comparison, a set of equivalent values is used. WI X3D AS is set at 0.5 degrees, MW SBR AS is set as Medium, and Sionna Fibonacci NS is set to $1.6e5$. This creates an equivalent starting condition. Fig. \ref{fig: Comparison simulations} illustrates the results, showing in each graph a different scenario and a wireless link simulation.

\begin{figure*} 
    
    \centering
    %\hspace{-1cm}
    %\subfloat[Average]{\includegraphics[height=4cm]{assets/radio_map/average_radiomap.png}\label{sufloat: radio_avg}}
    \subfloat[WI X3D]{
        \includegraphics[height=5cm, width = 0.28\linewidth]{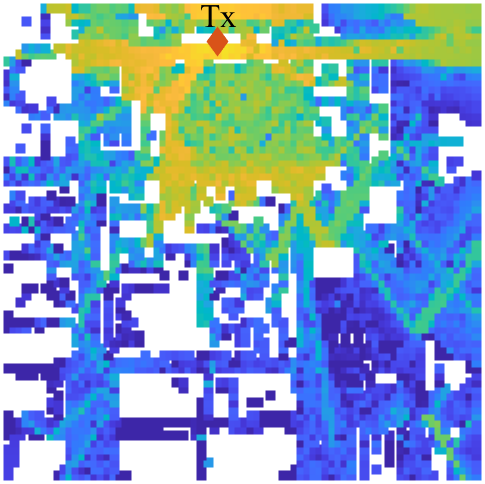}\label{subfloat: radio_wi}
    }
    \subfloat[Sionna Fibonacci]{
        \includegraphics[height=5cm, width = 0.28\linewidth]{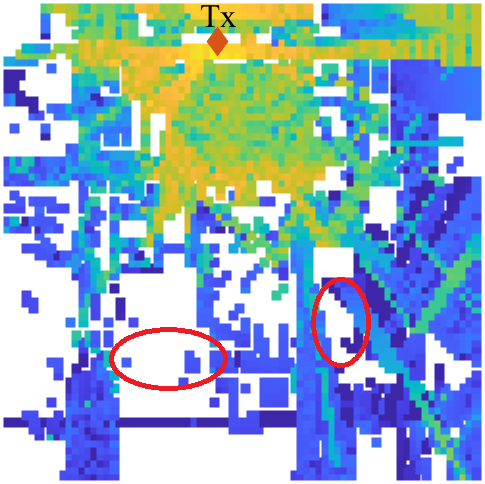}\label{subfloat: radio_sionna}
    }    
    \subfloat[MW SBR]{\includegraphics[height=5cm, width = 0.28\linewidth]{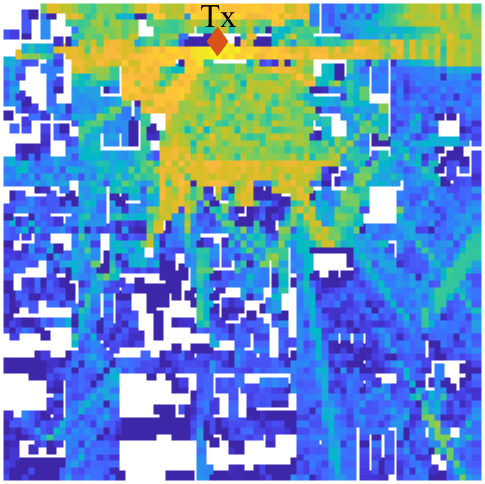}\label{subfloat: radio_matlab}}  
    \subfloat{\includegraphics[height=5cm]{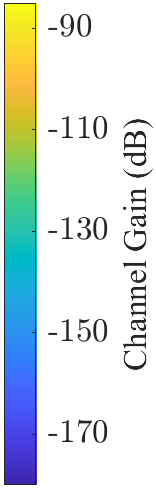}}

    \caption{Channel Gain (dB) of the radio map produced using the evaluated ray-based simulation software. Simulations are performed for a) WI X3D AS = 0.5 deg, b) Sionna Fibonacci NS = 1.6e5, c) MW SBR AS = Medium. WI and MW with maximum 3 reflections and 1 diffraction. Sionna maximum path depth 4. The Tx position is highlighted by a red diamond. Red circles in b) highlight areas that are reached by a combination of reflection and diffraction bounces, which the evaluated Sionna Fibonacci version is unable to model.}
    \label{fig: radio_map}
\end{figure*}

Comparison of Sionna in Fig. \ref{fig: comparison-sequential-scenario1} and Fig. \ref{fig: comparison-parallel-scenario1} reveals distinct behaviors in Sionna's computational time when utilizing CPU versus GPU across increasing interaction depth. The CPU behavior exhibits a linear trend with increasing path depth, whereas the GPU is not affected by path depth.
In Fig. \ref{fig: comparison-parallel-scenario1} Sionna GPU for maximum interaction depth of 1 observes a slightly higher computational time than the counterpart at depth of 5. This is due to the GPU warm-up time and just-in-time compilation overheads. 
%\footnote{A more accurate analysis would have required an higher number Monte Carlo simulations for the \textit{Multiple Link simulations}. However, since the paper aims at showing the overall behaviour of the simulation tools, was deemed unnecessary.}

Sionna Fibonacci supports the addition of diffuse scattering interactions without significantly increasing the average wall-clock time. The current model of Sionna Fibonacci does not model propagation paths that combine reflections and diffractions bounces. When diffuse scattering occurs, the Sionna model spawns a single additional propagation path. This approximation is valid if enough number of rays are spawned initially at Tx. 
Each diffuse scattering point for WI X3D is a new ray emission source, and the SBR algorithm is reintroduced spawning multiple propagation paths. This approach is more accurate, but introduces significant overhead in the computation.

%SL and ML simulations are evaluated at equivalent initial number of sampled rays---i.e., each extra tick signifies a better relative performance for given wireless link simulation without diffuse scattering. For simulations with diffuse scattering, the approximate range of increase is provided, measured at RLP of 0.5 deg and increasing maximum interaction depth.
%The behaviour at increasing maximum interaction depth is approximately linear for most of the simulations. 
%Sionna and WI implements different models for the diffuse scattering interaction, and its observed by their distinct behaviour in computational time.
%WI computes the exact interaction between the ray and encountered objects, while Sionna probabilistically models this interaction.

%Comparison of the selected ray-based simulation software over a comparable initial number of rays at Tx. WI X3D AS $=0.5$ deg, MW SBR AS = Medium, Sionna Fibonacci NS $=1.6e5$

\subsection{Radio Map predictions}
The three different simulation software are used to calculate the Channel Gain in Scenario2, obtaining the radio maps reported in Fig. \ref{fig: radio_map}.
These simulations were conducted with the following parameters: for WI X3D, an angular separation (AS) of 0.5 degrees; for MW SBR AS = Medium; and for Sionna, a Fibonacci number of samples (NS) of 1.6e5. WI and MW were configured with a maximum of three reflections and one diffraction, while Sionna was set to a maximum path depth of four.
The provided radio maps are visual indicators of the quality of the evaluated ray-based propagation models. They do not substitute comparisons against measurements and accurate coverage map prediction.

A single transmitter (Tx) was considered and is located on the top of the main building of the High Frequency Campus in Milan, Italy at a height of 21.7 m from the ground. The receivers are uniformly distributed throughout the map area at a height of 1.5 m, forming a grid with a resolution of 10 × 10 meters.
The carrier frequency is set to 28 GHz, and the Tx and Rx antennas are isotropic.

% \color{red}
% \sout{Overall, the results are consistent across the three simulators, with differences of only a few dB at specific points. WI X3D and MW SBR offer more detailed insights into the EM radio map, particularly where the Rxs are located far from the Tx. The impact of buildings and obstacles is represented more consistently by these two simulators compared to the Sionna Fibonacci.

% The WI X3D, shown in Fig. \ref{subfloat: radio_wi}, integrates the SBR method with a path correction algorithm that adjusts the interaction points to achieve precise propagation paths between the Tx and each Rx. This enhancement is particularly evident in non-line-of-sight points.

% Sionna Fibonacci, illustrated in Fig. \ref{subfloat: radio_sionna}, lacks channel characterization for certain RX points on the map. This limitation may be due to the underlying Mitsuba differentiable rendering system that is unable to compute the propagation path and/or due to the absence of path correction methods for Sionna Fibonacci.

% In Fig. \ref{subfloat: radio_matlab}, the radio map predicted by the MW SBR is shown. MW software also utilizes an SBR implementation with exact path correction, resulting in a comprehensive characterization of the EM environment.}

In general, the results are consistent between the three simulators. The computed channel gain differs only by a few dBs. The empirical distribution of the difference between the calculated channel gain and the average channel gain of the three simulators has been calculated, showing that no significant differences can be observed. This distribution has been omitted in this paper.

WI X3D and MW SBR, in Figs. \ref{subfloat: radio_wi} and \ref{subfloat: radio_matlab} respectively, offer more detailed insights into the radio map.
The Sionna Fibonacci radio map presents shadow areas, such as those highlighted with red circles in Fig. \ref{subfloat: radio_sionna}. In the other two simulators, those areas are reached by propagation paths that contain reflection bounces and diffraction bounces. The current implementation of the Sionna Fibonacci method cannot model propagation paths that mix reflection bounces and diffraction bounces. 
Sionna RT is a novel ray-based propagation simulation solution based on a differentiable rendering system. It is licensed under Apache 2.0 license, enabling the community to freely access the source code. The open-source approach is vital for fostering innovation, as it allows third parties to thoroughly evaluate and enhance the model.

\begin{table*} [!t]
	\centering
 \caption{%Summary of comparative results with \textit{ray launching parameter} (RLP) set to 0.5 deg, 
    Summary of results over a comparable initial number of rays at Tx. WI X3D AS $=0.5$ deg, MW SBR AS = Medium, Sionna Fibonacci NS $=1.6e5$.
    Single Link and Multi Link simulation performance, their approximate behavior at increasing maximum interaction depth, and the increase range with respect to the same simulation condition with the diffuse scattering interaction. Each extra tick indicates a better relative wall-clock time in the simulations without diffuse scattering.
    %and approximate behavior when diffuse scattering is considered. The approximate behavior with increasing max. interaction depth are summarised from the numerical results.
    }  
	\begin{tabular}{r|l |c| c| c| c}
		\toprule
        \multirow{2}{*}{\textbf{Simulation}} & \textbf{Software} & \textbf{Single Link} & \textbf{Multiple Link} & \textbf{Path Depth Behaviour} & \textbf{DS Behavior}  \\
        & & \textbf{Performance} & \textbf{Performance} & \textbf{Single/Multiple Link} & \textbf{Single/Multiple Link} \\
		% \textbf{Simulation} & \textbf{Software} & \textbf{Single Link simulations} & \textbf{Multiple Link simulations} & \textbf{DS Behaviour (Single/Multiple Link)} & \textbf{Path Depth Behaviour (Single/Multiple Link)} \\ %[2 pt]
		\noalign{\smallskip}
		\hline
		\noalign{\smallskip}

        % OJCOMS guideline: to indicate range of values write "7 to 9"
        \multirow{2}{*}{Scenario1 }
        & Sionna Fibonacci & \checkmark\checkmark & \checkmark\checkmark & linear, constant & $\times$(1.5 to 2), constant  \\
        & WI X3D & \checkmark & \checkmark & linear, linear & $\times$(1 to 1.5), $\times$(10 to 25) \\ 
        \midrule

        \multirow{3}{*}{Scenario2 }
        & Sionna Fibonacci & \checkmark\checkmark\checkmark & \checkmark\checkmark & linear, constant & $\times$(1.5 to 2), constant \\
        & WI X3D & \checkmark\checkmark & \checkmark\checkmark\checkmark & linear, linear & $\times$(1 to 1.5), $\times$(20 to 30) \\ 
        & MW SBR & \checkmark & \checkmark & linear, linear & N/A \\

		\bottomrule
	\end{tabular}

	\label{tab: rt overview}
\end{table*}

\section{Overview and Discussion} \label{sec: discussions}
This section aims to summarize the main observations and the notable points derived from our simulations.
Table \ref{tab: rt overview} provides an overview of the results evaluated with an equivalent initial number of rays. Each extra tick in the SL and ML performance column indicates a better relative wall-clock time for the given wireless link simulation without diffuse scattering. With an increase in the maximum interaction depth, the behavior is approximately linear for most of the simulations. The diffuse scattering behavior indicates the approximate increase range with respect to the same simulation parameter without the diffuse scattering interaction. Sionna and WI implement different models for the diffuse scattering interaction, as can be observed by their distinct behavior. 

The computational performance for ray-based models is a multifaceted task. This includes the meshes of the propagation environment, the underlying computing infrastructure, and any type of algorithmic optimization. The type and number of interactions also significantly affect the simulation times. Conducting a comprehensive analysis from a computational theory standpoint of all these factors is challenging and becomes unfeasible when the source code is unavailable.
Instead, benchmarks serve as a practical alternative for evaluating algorithmic implementations and hardware configurations in a black-box manner. 

The simulations that compare the radio map using three different software in similar simulation settings showcase consistent results across them with minor discrepancies. WI X3D and MW SBR provided more detailed information on the EM radio map. 
% \textcolor{red}{\sout{particularly for distant receiver points, while Sionna Fibonacci lacks characterization for certain Rx locations in non-line-of-sight condition.} % due to its balancing of accuracy and computational efficiency. 
The evaluated Sionna Fibonacci version does not model propagation paths composed of a combination of reflection bounces and diffraction bounces.

\section{Conclusions} \label{sec:conclusion}

This paper proposes a benchmark of different ray launching solutions, considering several simulation parameters such as the angular separation/number of samples and ray interactions with the environments. Two different urban environments are considered, one with complex mesh detail and one with simpler mesh detail. Two sets of transceiver configurations that evaluate different use cases are tested.
The proposed framework allows the practitioner to make an informed decision when selecting the appropriate solution for their application. The availability and reproducibility of the benchmark\footnote{Source code and scenarios are available in https://github.com/Michele-Zhu/ray-launching-benchmark.} have the additional advantage of providing a reference point to track advances, ensuring that improvements in computational performance can be systematically monitored and evaluated.

\bibliographystyle{IEEEtran}
\bibliography{Bibliography}

\vskip -1\baselineskip plus -1fil

\begin{IEEEbiography}[{\includegraphics[width=1in,height=1.25in,clip,keepaspectratio]{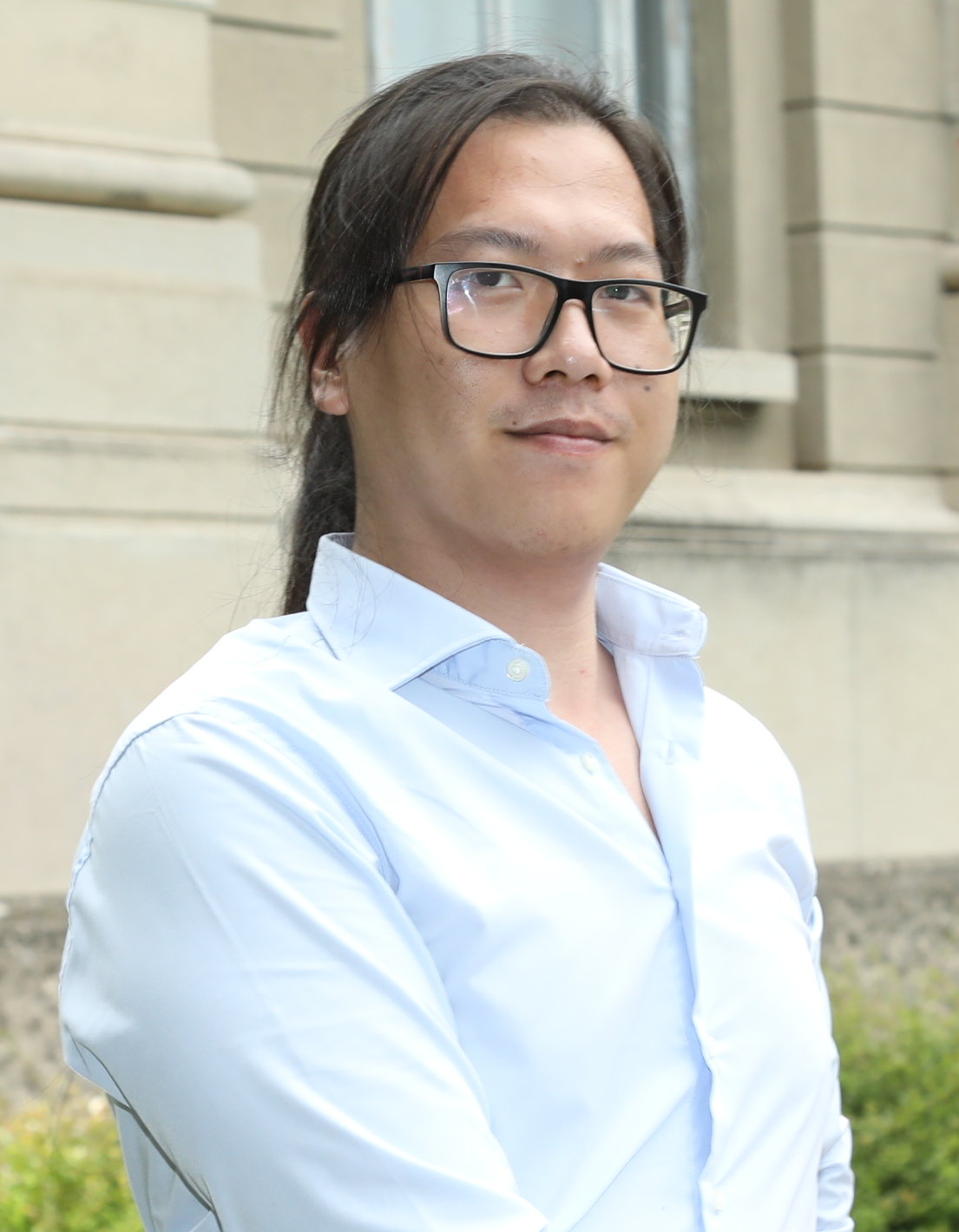}}]{Michele Zhu } (Graduate Student Member, IEEE) received the B.Sc. in information engineering from Università di Padova, Padua, Italy, and the M.Sc. degree in computer science and engineering from Politecnico di Milano, Milan, Italy. He started pursuing the Ph.D. degree in information technology in September 2023. He is involved in research projects in the Joint Laboratory between Huawei Technologies Italia, Milan, and Politecnico di Milano. His research interest are on deep learning with applications in wireless communication systems and semantic communication.
\end{IEEEbiography}

\vskip -1\baselineskip plus -1fil

\begin{IEEEbiography}[{\includegraphics[width=1in,height=1.25in,clip,keepaspectratio]{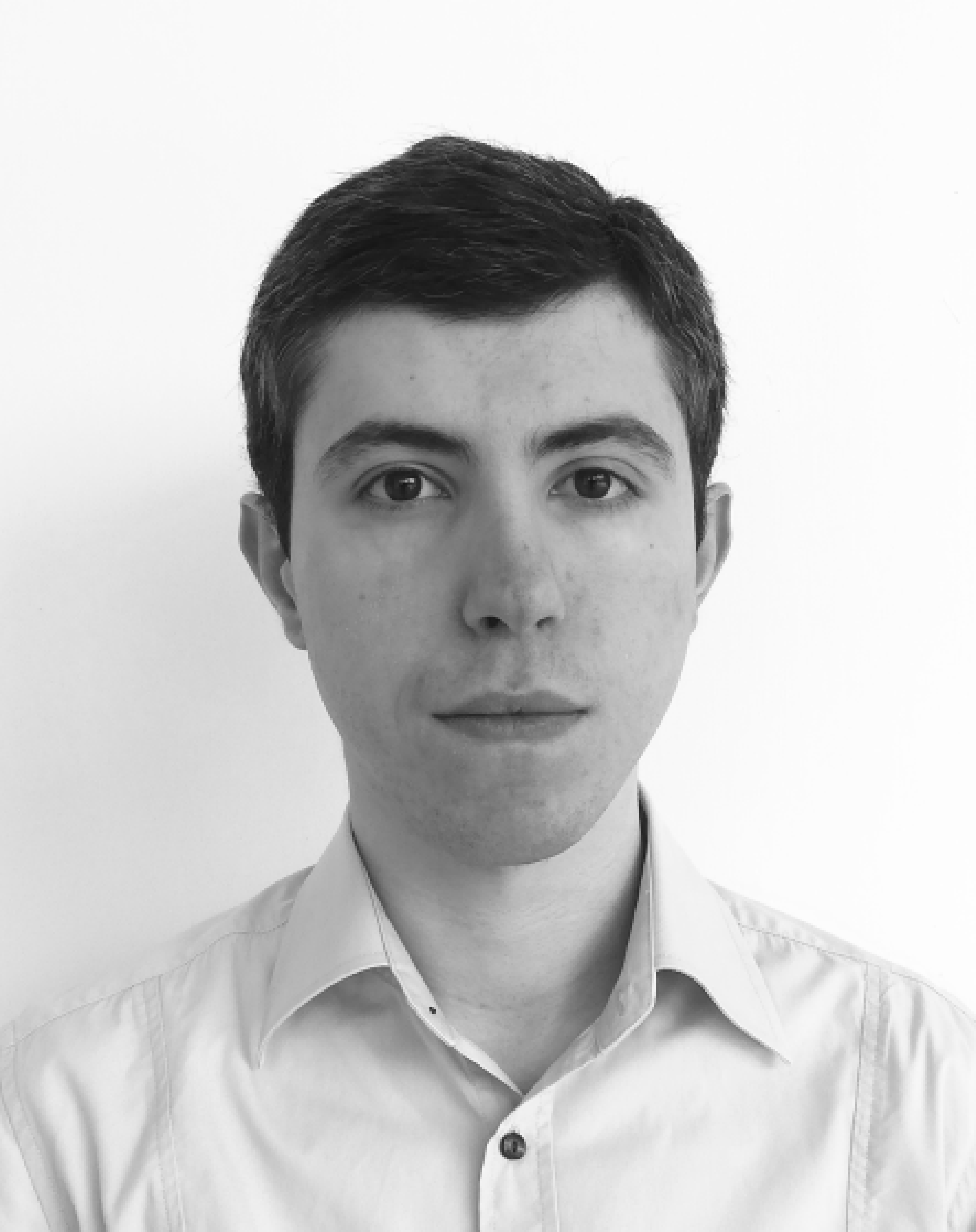}}]{Lorenzo Cazzella } (Graduate Student Member, IEEE) received the B.Sc. degree in information engineering from Università del Salento, Lecce, Italy, in 2017, and the M.Sc. degree in computer science and engineering from Politecnico di Milano, Milan, Italy, in 2020, where he is currently pursuing the Ph.D. degree in information technology. During his master thesis, he has been involved in the Joint Laboratory between Huawei Technologies Italia, Milan, and Politecnico di Milano, where his research work nowadays concerns the study and development of deep learning solutions for channel estimation and integrated communication and radar sensing at the infrastructure. His research interests are currently focused on geometric deep learning and unsupervised learning techniques, with applications to wireless communication systems and radar systems.
\end{IEEEbiography}

\vskip -1\baselineskip plus -1fil

\begin{IEEEbiography}[{\includegraphics[width=1in,height=1.25in,clip,keepaspectratio]{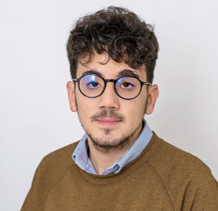}}]{Francesco Linsalata  }(Member, IEEE) received M.Sc and PhD degrees cum laude in Telecommunication engineering from Politecnico di Milano, Milan, Italy, in 2019 and 2022, respectively. He is a researcher at the Dipartimento di Elettronica, Informazione e Bioingegneria, Politecnico di Milano. His main research interests focus on V2X and UAV communications, Integrated Communication and Sensing, and physical layer design for 6G wireless networks. He was co-recipient of one best-paper award and recipient of one best student paper award.
\end{IEEEbiography}

\vskip -1\baselineskip plus -1fil

\begin{IEEEbiography}[{\includegraphics[width=1in,height=1.25in,clip,keepaspectratio]{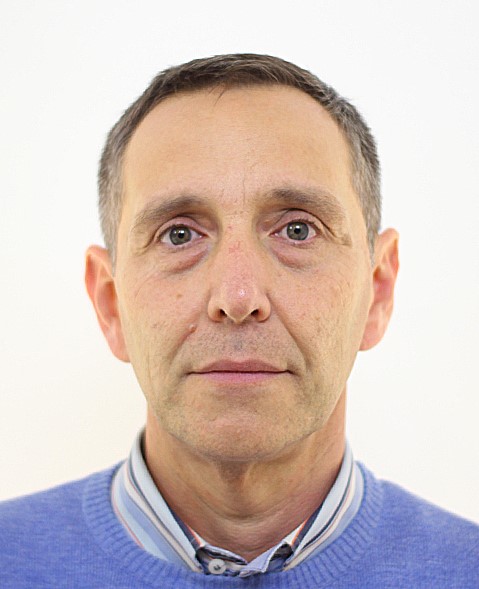}}]{Maurizio Magarini }  (Member, IEEE) received the M.Sc. and Ph.D. degrees in electronic engineering from Politecnico di Milano, Milan, Italy, in 1994 and 1999, respectively. In 1994, he was granted the TELECOM Italia Scholarship Award for the M.Sc. Thesis. He was a Research Associate with Dipartimento di Elettronica, Informazione e Bioingegneria, Politecnico di Milano, from 1999 to 2001. From 2001 to 2018, he was an Assistant Professor with Politecnico di Milano, where he has been an Associate Professor since June 2018. From August 2008 to January 2009, he spent a sabbatical leave with Bell Labs, Alcatel-Lucent, Holmdel, NJ, USA. He has been involved in several European and National research projects. His most recent research activities have focused on molecular communications, massive MIMO, vehicular communications, wireless sensor networks for mission critical applications, study of advanced waveforms for 5GB and 6G, and wireless networks using unmanned aerial vehicles and high-altitude platforms. He has authored or coauthored more than 180 journal articles and conference papers. His research interests include communication and information theory, synchronization, channel estimation, equalization, and coding applied to wireless and optical communication systems. He was a co-recipient of four best-paper awards. He is an Associate Editor of IEEE Access and IET Electronics Letters. He is an Area Editor of Nano Communication Networks (Elsevier).
\end{IEEEbiography}

\vskip -1\baselineskip plus -1fil

\begin{IEEEbiography}[{\includegraphics[width=1in,height=1.25in,clip,keepaspectratio]{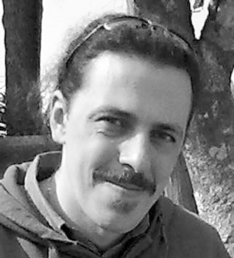}}]{Matteo Matteucci } (Member, IEEE) received the M.Sc. degree in knowledge discovery and data mining from Carnegie Mellon University, and the Ph.D. degree in computer engineering and automation from Politecnico di Milano. He is a Full Professor with the Dipartimento di Elettronica Informazione e Bioingegneria of Politecnico di Milano, Italy. His research interests are mainly in pattern recognition, robotics, computer vision, and signal processing. He has co-authored more than 50 (peer-reviewed) papers on international journals, 25 papers in international books, and more than 150 (peer-reviewed) contributions to international conferences and workshops.
\end{IEEEbiography}

\vskip -1\baselineskip plus -1fil

\begin{IEEEbiography}[{\includegraphics[width=1in,height=1.25in,clip,keepaspectratio]{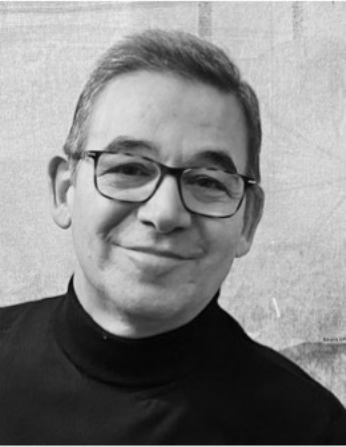}}]{Umberto Spagnolini } (Senior Member, IEEE) is a Professor of Statistical Signal Processing, the Director of Joint Lab Huawei-Politecnico di Milano, and the Huawei Industry Chair of the Politecnico di Milano. His research in statistical signal processing covers remote sensing and communication systems with more than 380 papers on peer-reviewed journals/conferences and patents. He is the author of the book Statistical Signal Processing in Engineering (J. Wiley, 2017). He is the Technical Expert of standard-essential patents and IP. His research interests include mmW channel estimation and space-time processing for single/multiuser wireless communication systems, cooperative and distributed inference methods, including V2X systems, mmWave communication systems, parameter estimation/tracking, focusing and wavefield interpolation for remote sensing (UWB radar and oil exploration), and integrated communication and sensing. He was the recipient/co-recipient of the Best Paper Awards on Geophysical Signal Processing Methods (from EAGE), the Array Processing (ICASSP 2006), the Distributed Synchronization for Wireless Sensor Networks (SPAWC 2007, WRECOM 2007), the 6G Joint Communication and Sensing (JC\&S 2021), and the SAR Imaging for Automotive (MMS 2022). He served as a part of IEEE editorial boards as well as a member in technical program committees of several conferences for all the areas of interests.
\end{IEEEbiography}

\end{document}